\documentclass[twocolumn]{article}
\usepackage[latin9]{inputenc}
\usepackage{graphicx}
\usepackage{url}
\usepackage[authoryear]{natbib}

\usepackage{verbatim}
\usepackage{caption}
\usepackage{wrapfig}
\usepackage{algorithm}
\usepackage{algorithmic}

\usepackage{xcolor}

\newcommand\blnote[1]{%
  \begingroup
  \renewcommand\thefootnote{}\footnote{\kern-5pt \textcolor{white}{\rule{5pt}{2ex}}#1}%
  \addtocounter{footnote}{-1}%
  \endgroup
}

\usepackage{fancyhdr}
\fancypagestyle{plain}{%
  \fancyhf{}
  \fancyfoot[C]{\thepage\\[2.5ex]{\footnotesize Distribution Statement ``A'' (Approved for Public Release, Distribution Unlimited)}\\[0.8ex]{\footnotesize \copyright BAE Systems 2021 all rights reserved}\\[0.8ex]{\footnotesize Not export controlled per ES-FL-042721-0053}}
  
  }
\begin{document}
\author{Bernard McShea$^1$, Kevin Wright$^1$, Denley Lam$^1$, Steve Schmidt$^1$,\\ Anna Choromanska$^{2,\ddagger}$,  Devansh Bisla$^{2,\ddagger}$, Shihong Fang$^{2,\ddagger}$,\\ Alireza Sarmadi$^{2,\ddagger}$, Prashanth Krishnamurthy$^{2,\ddagger}$, Farshad Khorrami$^{2,\ddagger}$%
  \blnote{$^1$ FAST Labs, BAE Systems, Arlington, VA 22209, USA.}%
  \blnote{$^2$ Dept. of Electrical and Computer Engg., NYU Tandon School of Engg., Brooklyn, NY 11201, USA.}%
  \blnote{$^\ddagger$ Equal contribution.}
}
\title{ESAFE: Enterprise Security and Forensics at Scale  \thanks{This work was supported in part by DARPA under Space and Naval Warfare Systems Center, Pacific (SSC Pacific) contract N66001-18-C-4037. The views, opinions and/or findings expressed are those of the author and should not be interpreted as representing the official views or policies of the Department of Defense or the U.S. Government.}}
\date{March 1, 2021}

\maketitle 

\begin{abstract}
Securing enterprise networks presents challenges in terms of both their
size and distributed structure. Data required to detect and characterize
malicious activities may be diffused and may be located across network
and endpoint devices. Further, cyber-relevant data routinely exceeds
total available storage, bandwidth, and analysis capability, often
by several orders of magnitude. Real-time detection of threats within
or across very large enterprise networks is not simply an issue of
scale, but also a challenge due to the variable nature of malicious
activities and their presentations. The system seeks to develop a
hierarchy of cyber reasoning layers to detect malicious behavior,
characterize novel attack vectors and present an analyst with a contextualized
human-readable output from a series of machine learning models. We
developed machine learning algorithms for scalable throughput and
improved recall for our Multi-Resolution Joint Optimization for Enterprise
Security and Forensics (ESAFE) solution. This Paper will provide an
overview of ESAFE's Machine Learning Modules, Attack Ontologies,
and Automated Smart Alert generation which provide multi-layer reasoning
over cross-correlated sensors for analyst consumption. 
\end{abstract}

\section{Introduction}

While current-day enterprise networks are becoming increasingly larger
and more complex, cyber-adversaries are also becoming increasingly
more sophisticated with an ever enlarging repertoire of attack tools
and techniques. New attack vectors are fueled by vulnerabilities of
various kinds in increasingly more complex software and made more
dangerous with increasing connectivity and scale of enterprise networks.
Cyber-security solutions are therefore increasingly crucial and challenging.
Robust solutions requires a multi-layer approach that brings together
multiple sensors from a variety of vantage points (e.g., network logs,
kernel and application logs, etc.) and performs robust automated threat
detection from these sensor feeds. Key requirements for a robust cyber-security
solution include: 
\begin{itemize}
\item Threat detection at scale: With current-day enterprise network sizes,
the various sensor logs discussed above are truly massive in volume.
To be able to process data in near-real-time, an effective threat
detection system must be able to scale to very large data volumes. 
\item Support for multiple types of sensor data: The various sensors discussed
above provide data with very different types of semantics, different
time scales, and different structures of numerical and categorical
data. An effective threat detection system must be able to scale to
different types of sensor data using a coherent unified algorithmic
core. 
\item High accuracy, precision, and recall: Due to the massive data volumes,
very high accuracy is crucial for threat detection systems to be effective.
False positives (i.e., spuriously declaring threats) are undesirable
since any such spurious threat detection would need to be processed/triaged
in some way (e.g., by a human operator) and even a low percentage
of false positives can quickly overwhelm such a processing/triaging
component. When confronted with the possibility of a significant number
of false positives, a human operator would, for example, start ignoring
the threat detections by the system, thereby making the detection
system ineffective. 
\item Human-understandability of underlying logic/reasons for threat detections:
To be able to interpret threat detections and determine possible responses,
a human analyst requires a human-comprehensible context for the threat
detections and the reasons for why a particular set of data points
(e.g., network log entries) were considered a threat. Hence, to be
truly effective, an automated threat detection system must be able
to provide a human-interpretable identification of the particular
fields/features that triggered a threat detection and the decision
logic based on these identified fields/features that informed the
alert. 
\end{itemize}
To address the challenges highlighted above, we propose Multi-Resolution
Joint Optimization for Enterprise Security and Forensics (ESAFE),
an integrated framework that brings together multiple complementary
ingredients to achieve a robust, scalable, and high-accuracy cyber-security
solution: 
\begin{itemize}
\item To leverage the understanding of signatures of ``localized'' elemental
events to enable detection of novel attacks, ESAFE's operation is
defined in terms of sequences/clusters of elemental labels. Each elemental
label is indicative of specific events or behaviors (e.g., failed
logons, port scans) and sequences/clusters of such labels are defined
as indicative of larger-scale malicious/anomalous behavior. 
\item For efficient at-scale classification of network/host-based data for
detection of elemental labels, a deep neural network (DNN) based methodology
is applied based on the Gated Mixture of Experts (GME) architecture.
This architecture is scalable to a wide variety of data schemas and
can operate at scale to fuse data from multiple sources while automatically
picking the specific data source that is most informative for each
type of label being considered in multi-label classification. 
\item To extract a human-understandable context for event classifications
and threat detections, a probabilistic parameterized Metric Temporal
Logic (ppMTL) based approach is applied, which identifies the specific
combinations of features from the sensor data that informed each specific
threat detection. The ppMTL component identifies which features are
relevant for threat detection and a human-readable representation
of the decision logic in terms of these features (as a set of linear
inequalities among these identified features). 
\end{itemize}
This paper is organized as follows. The methodology for generation
of elemental labels and clusters thereof is discussed in Section~\ref{sec:labeling}.
The machine learning methodologies based on GME and ppMTL approaches
outlined above are described in Section~\ref{sec:ML}. The overall
system architecture is discussed in Section~\ref{sec:system_architecture}.
The lessons learned are summarized in Section~\ref{sec:lessons_learned}.
The open challenges and future work are discussed in Section~\ref{sec:open_challenges}.

\section{Labeling Methodology}

\label{sec:labeling}

The critical first step in the application of supervised machine learning
with respect to the cyber security domain, in particular to the task
of finding unknown intrusions via log data, is labeling the data.
This presents several nontrivial challenges including: highly unbalanced
dataset, intrinsic difficulty in defining what constitutes a network
intrusion (which is still an open problem), and labeling ``normal''
data given the high noise floor in network traffic. Prior research
(Sommer and Paxson \citet{MachineLearning}) describes the difficulty in observing
a sufficient amount of network traffic for enabling classification
to have adequate accuracy.

Our first approach was to generalize the current best practice of
known signatures and heuristics rules. In generalizing the current
rule base, we gained the ability to detect similar behavior at the
expense of incurred nominal higher false positive rate for individual
rules. Our intention was that secondary reasoning over the initial
labeling of the data would consider multiple data sources over a given
finite window of time, thus forming identification of a threat vector.
This enabled us to summarize for a given system which of our rules
fired during a given time window. While this is a minor advancement
in the current state of the art in terms of a condensed representation
of current generalized rules, it does not address discovering unknown/unseen
attacks. Since the rules are directly based on current Security Operations Center (SOC) rules, it
bridges the \emph{semantic gap} as defined by Sommer and Paxson (\citet{MachineLearning})
to the current state of practice. They define the semantic gap as
effort needed to transfer results into actionable reports for the
network operator. Another issue with this approach is scaling: the
rate at which new rules can be written can only match current industry
standards and these limitations lead us to consider a constructive
approach.

Our following approach for labeling data was to deconstruct attack
tactics techniques procedures (TTPs) to network artifacts and further
deconstruct network artifacts to elemental labels on particular logs.
This allowed ESAFE to construct elemental labels from known attacks
and form clusters of elemental labels to infer particular TTPs. Below,
we discuss two approaches for constructing elemental labels and clusters.
In combination, these clusters provide five distinct benefits. 
\begin{enumerate}
\item Shifts in TTPs from attackers will continue to trigger most of the
elemental labels since attackers often evolve their techniques. 
\item Unknown/unseen TTPs can be discovered as hybrids of current known
TTPs through identification of appropriate composite clusters. 
\item Elemental labels are less likely to have legacy issues as they cover
a dynamic behavior observed in a log rather than hard-coded rules.
As an example of a legacy issue, proxy rules that block a particular
domain ``www{[}dot{]}mal{[}dot{]}com'' will not fire once the attacker
changes the domain. 
\item Logs can be studied and elemental rules can be written to ensure coverage
prior to having seen evidence of their use by any TTP. 
\item Adding a single elemental label can inform many other clusters. This
provides a non-linear way to scale the production of clusters which
can be detected. 
\end{enumerate}

\subsection{Data Driven Elemental Labels}

Our bottom up approach was to analyze individual logs in order to
create data driven elemental building blocks. We created elemental
rules for: single rows of data, multiple fixed length rows, and rolling
time windows. An example of an elemental rolling time window rule
is given by the label ``Abandoned Logon Attempt'', wherein the corresponding
rule looks for two failed attempted logons and the absence of a successful
logon within a given window. Taken in isolation, this elemental feature
is not necessarily indicative of an attack and is strictly informative
when taken as part of a cluster comprising a network artifact. This
elemental feature could be, for example, simply an occurrence of a
single user not remembering their password and stopping short of locking
their account, supporting evidence of a brute force attack on the
network, or many other possibilities. An example of a single row event
would be a Windows Event Log ID {[}4719, 4715, 4812, or 4885{]} indicating
that there was a change in the system audit policy; again, we see
an elemental feature describing what is happening, but which by itself
does not indicate an attack. Some elemental features, "anomalous
request type cipher for server", are derived directly from statistical
studies of the data. This rule fires for a given server IP if the
(version, cipher) pair is anomalous.

\subsection{Data Driven Clusters}

Our bottom up approach to clustering, taking elemental features and
reconstructing network artifacts, was to simply find all labels for
a given IP and create a weighted threat vector based on number and
severity of elemental features. We then sorted this list and designated
the top N nodes as starting points. From our starting points, we map
out interactions (IP connections) to all other nodes from our candidate
node and repeat the ranking process and follow the next top second
degree N nodes. We repeated this process for a max depth of three
hops. This was a manual process, which could be automated. The greater
concern here is interpreting the cluster with respect to the \emph{semantic
gap}.

Data Driven Elemental Labels (DDEL) provides a method for labeling
cyber data without the expectation or knowledge of TTPs. This technique
is critical to answering the question: how do we find unknown intrusions
via log data? However, the Data Driven Cluster is difficult and in
some cases impossible to interpret by the cyber analyst. Exclusively
clustering DDEL did not decrease the \emph{semantic gap}. This motivated
the approach to ontological clustering described below.

\subsection{Attack Ontology Cluster}

Our top down approach is to label known clusters by constructing elemental
labels with the available supporting data of known artifacts. Let
us indulge in the medical analogy of symptoms and diagnoses. The top
down approach utilizes predefined attack types (illnesses) for which
elemental labels (symptoms) would eventually through the series of
algorithms, described in the following sections, have the ability
to be aggregated to support known suspicious activity. For example,
given a cough, sore throat, and fever, an inference of the flu would
be an actionable assessment of the symptoms. This methodology can
then easily be extended to clusters or named attacks (illnesses) which
are modified in small ways to attempt to avoid detection. A modified
cluster would be considered highly suspicious if it contained many
elemental indicators to a known attack cluster and exhibited slight
modifications. Also, a hybrid of two known techniques would be detected
as one new cluster with components of two partial clusters and considered
suspicious. Attack ontology elemental labels create an extensible
foundation for detecting named attacks. The \emph{semantic gap} is
bridged by domain knowledge in this approach with the limitation that
highly novel attacks, clusters of labels with few or no elemental
labels in known clusters, will be missed. 
\begin{figure}[!h]
\includegraphics[width=1\linewidth]{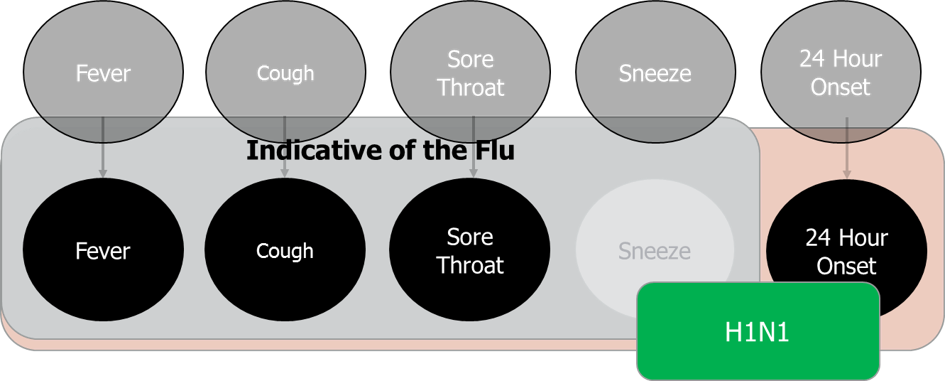} \caption{Analogy To Virus Indicators}
\label{fig:H1N1} 
\end{figure}

\begin{figure}[h!]
\includegraphics[width=1\linewidth]{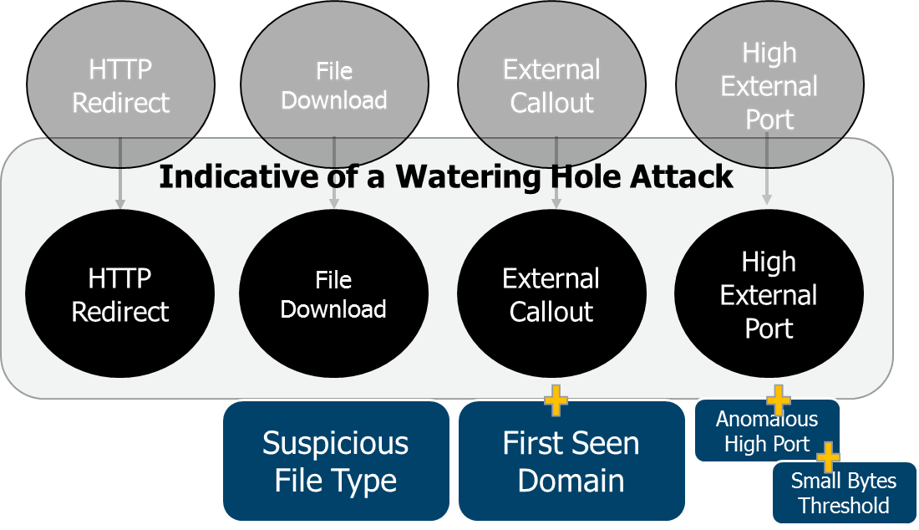} \caption{Cluster Ontology}
\label{fig:Cluster Ontology with Enrichments} 
\end{figure}

\subsection{Attack Ontology Elemental Labels}

Once a class of malicious behavior to search for has been identified,
the log artifacts that such a behavior would leave behind need to
be identified and elemental labels of the smallest atomic units would
need to be defined. For example, a network connection, a file modification
log entry, or a JA3 hash are all Attack Ontology Elemental Labels
(AOEL). In our experiments, we attempted to identify 3-5 AOEL in at
least two different data types for each class of malicious behavior
we were hunting for. Although we were able to demonstrate some success
with this approach, we recognize the following open problems; determining
the optimal number of AOEL per cluster, maximizing the abstraction
of the attack ontology ensuring cluster labels have the correct overlap,
and ensuring that AOEL have coverage of all possible TTP artifacts
for a given label. 
\begin{figure}[h!]
\includegraphics[width=1\linewidth]{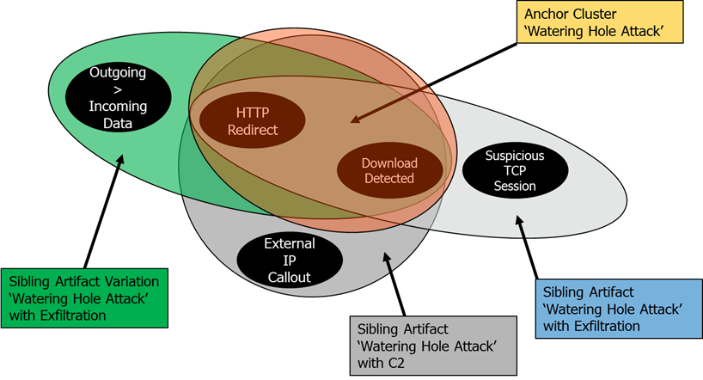} \caption{Attack Ontology Clustering}
\label{fig:Attack Ontology Clustering} 
\end{figure}

\subsection{Balanced Constructive Labeling Method}

In order to discover new unknown attacks and TTPs and bridge the \emph{semantic
gap}, there is no single one \emph{best} labeling strategy and we
need a moderated approach where we have enough AOEL rules to have
clusters which are explainable to the user and enough DDEL rules able
to find new attacks.

\section{Replacing Database Heuristics with Machine Learning}

\label{sec:ML} In order to scale to large volumes of data and future
perturbations of attacks, there must exist a methodology to appropriately
label or enrich the data early on in the reasoning system. The elemental
feature enrichments come in a variety of complexities as mentioned in
Section 2. Machine Learning classifiers vary in their use cases for
certain types of classification, however, in our Cyber Reasoning System,
we require partly feature detection and partly feature generation
(multilabel classes). While such a machine learning classification
can be addressed using a Convolutional Neural Network (CNN), training
a vanilla CNN on all concatenated data types for all potential label
classes will not work well for networks missing a sensor or experiencing
sensor failure. As well, not all labels are of the same complexity,
meaning classifier training on labels consisting of single feature
single value is different than training for a multiple feature multiple
value threshold. In order to enrich the alert feeds coming from commercial
and proprietary systems, we apply a Gated Mixture of Experts (GME)
architecture. The GME architecture below solves the scale and missing
inputs issues while allowing a probabilistic model to operate on the
input data. This in itself is fundamental to scalability of finding
novel attacks or small perturbations on known attacks. Since labels
(enrichments) are not hard coded, the decision boundary hyperplane
is in a way fuzzing the rule set. Labels involving strings can allow
for string fuzzing, thresholds such as number of bytes can be learned
and dynamic in nature from new input data, and series of nested AND/OR
statements in traditional named attacks can be reasoned or learned
later down the system processing pipeline via their elemental enrichments.
To provide a human-understandable explanation of the classifications
and resulting threat detections, we apply a probabilistic parameterized
Metric Temporal Logic (ppMTL) based approach. The ppMTL based component
generates human-readable annotations that identify the specific features
in the input data that informed a particular classification and the
particular decision logic in terms of these identified features (as
a set of linear inequalities). To facilitate human-understandability
of the generated explanations, a crucial part of the ppMTL approach
is construction of spatio-temporal aggregations of input data (e.g.,
network logs) from the viewpoints of individual IP addresses in the
network. These spatio-temporal aggregations are constructed using
a dynamic graph methodology. The GME and ppMTL components in ESAFE
are described in Sections~\ref{sec:GME} and \ref{sec:ppmtl}, respectively.

\subsection{Gated Mixture of Experts}

\label{sec:GME} 
\begin{figure}[h!]
\includegraphics[width=1\linewidth]{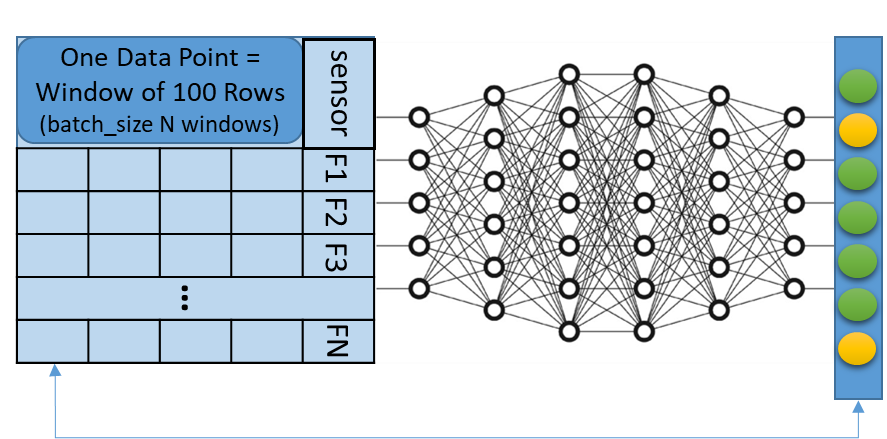} \caption{Single Expert}
\label{fig:Frozen Sensor Architecture} 
\end{figure}

\begin{figure}[h!]
\includegraphics[width=1\linewidth]{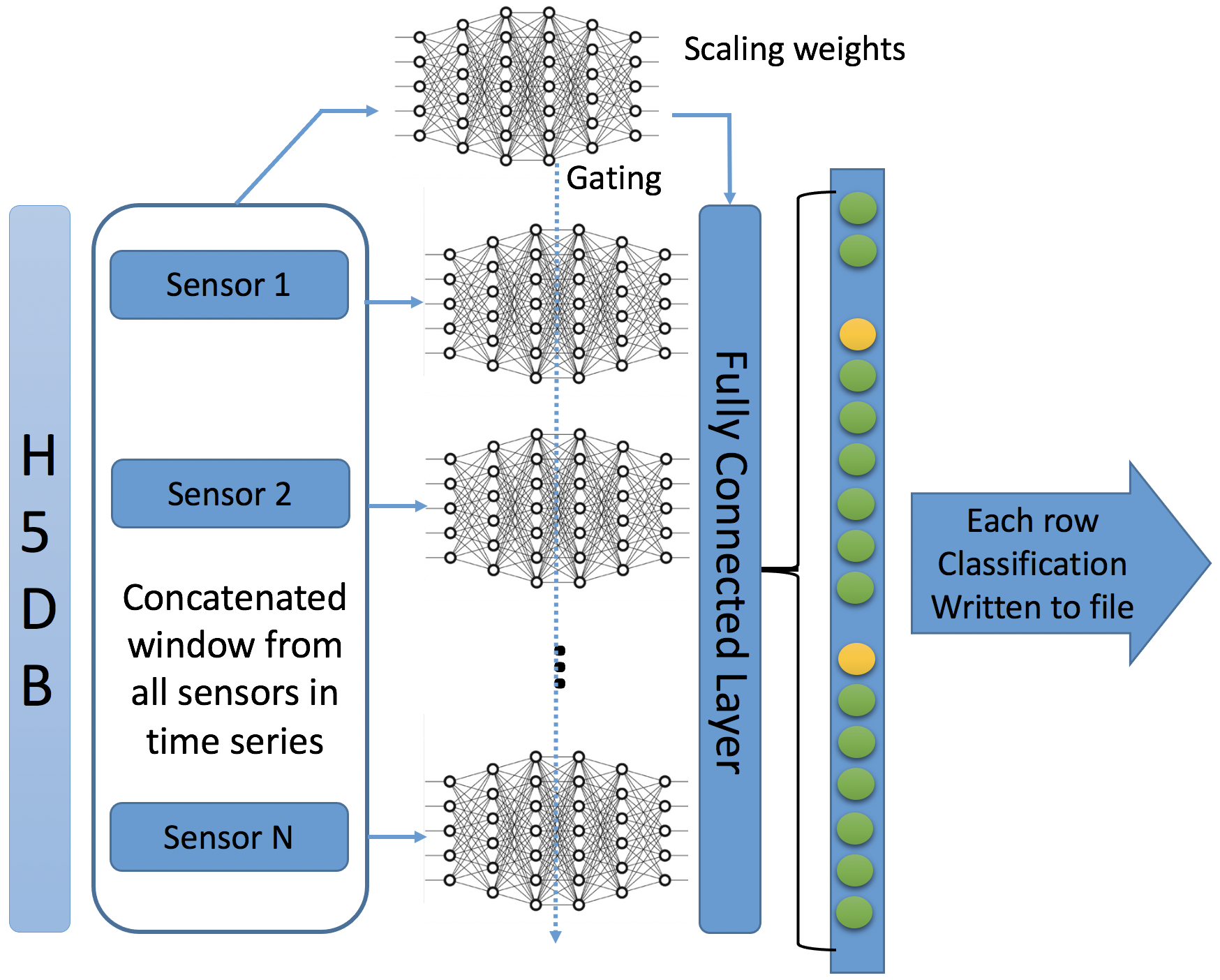} \caption{Gated Mixture of Experts Architecture}
\label{fig:Gated Mixture of Experts} 
\end{figure}

Crucial challenges in the development of a machine
learning based classifier for the cyber-security problem addressed
by ESAFE include: scalability to massive amounts of data (e.g., training
and running inference over terabytes of data), requirement for near-real-time
performance (since robust cyber-security requires rapid visibility
into on-going attacks; also, slower than real-time throughput will
result in growing back-logs), robustness to varying availabilities
of individual sensor feeds (e.g., absence of some sensors on some
networks or for some time periods) and ability to generate classification
outputs with only partially observable or missing inputs. The GME's
main function in the ESAFE pipeline is multilabel classification of
event indicators (labels discussed in Section~\ref{sec:labeling})
for each row of data fed in as part of a window of $N$ rows of data
(where $N$ is a window size parameter, e.g., $N=100$ is the default
setting in our experiments). The high level architecture is shown
in Figs.~\ref{fig:Frozen Sensor Architecture} and \ref{fig:Gated Mixture of Experts}.
The pipeline of data consists of fetching a window across sensors,
data formatting, and inference. The data from each sensor is processed
by a specialized network (an {\em expert}) for that particular
sensor data and the outputs of the different experts are combined
using a gating mechanism. 

\subsubsection{Frozen Single Experts}

Each individual expert is pretrained as a current golden model; this
allows for off-line training and competition of historical models
against newer data as the dynamic nature of the dataset could change
by the day or hour. These models comprise the initial layer of the
GME architecture. This is an important feature as it allows for specific
tuning and data segmentation based on the cyber-network being addressed,
volume of data, and data types. In our experiments, we researched
a variety of individual expert types, eventually settling on each
individual sensor being its own expert type. Previously, we had also
looked at dividing the data by subdomain, country, or protocol type,
and having separate experts for each such division. But, such finer-granularity
divisions were not found to provide significant benefits and the per-sensor
expert was found to provide the best speed and accuracy. 
Each of our labels is
based on data from a single sensor source; hence, per-sensor experts
provide optimal performance in our setting. 
However, it should be noted that labels based on multiple sensor sources
can be addressed in the system by segmenting the data appropriately
allowing for a more complex label to exist in a single expert segmentation.
Although the hyperparameters for each expert can be tuned separately,
it was found that only minimal such tuning is required and training
hyperparameters can be largely retained when moving to new sensor
data types. 

\subsubsection{Batch Retrieval}

Data for the GME was stored in a HDF5 database, which 
combines the sorted containers of data for each available sensor by
time. Each individual sensor's data is organized by time (minimally
error-prone using 1 minute files and general commercial output accuracy).
From these individual containers, a single sorted container is created
prior to the GME running training or inference. As batches of windows
are fetched across all sensors (or zeroed out), the individual expert
layers receive a comparable time window for classification into
the most probable classes. Batches are created with batch
size and window as parameters.

\subsubsection{Classification in the Gated Mixture of Experts}

Given multiple sensor data sources, the data from the sensors are
aligned over time windows as described above so that the window extracted
from each sensor feed provides a view into the cyber-network's operation
over the same time range. With multiple sensors, the per-sensor experts
operate as sensor-specific feature extractors, which are combined
together using a gating mechanism. For this purpose, a separate light-weight
gating network is trained that takes each sensor's data and generates
per-sensor scaling factors (which can be conceptually viewed as estimates
of the importance of each sensor for classification in that particular
time window). These scaling factors are then used to weight the outputs
of the individual experts (i.e., to automatically determine which
sensors are most relevant in that time window). The scaled combination
of features is then passed into a fully connected layer followed by
a softmax layer with binary cross entropy log loss function for training.
The gating mechanism provides robustness to missing sensors or missing
data in a time window. This is a crucial feature of the GME to allow
decision making under periods of unobservable behavior. 
Given a time window of data, GME's classification output for each
row of data is either ``Normal'' or one or more labels. As rows
of data are written out with label(s), this inherently allows for
the most important rows contributing to higher order named threat
classes to pass through. This was shown to filter the dataset from
over several terabytes to 100's of gigabytes while preserving a mapping
index into the full raw dataset. These classifications should be viewed
as enrichment data points versus malicious or even suspicious elements
on their own. The classification by a machine learning algorithm provides
an alternative to database heuristics as the machine learning adds
elements of both fuzzy matching (multi-field labels) and threshold
finding (continuous fields). The labels output by GME are cross-correlated
further down the data flow pipeline in ESAFE by the post-processing
module (including cross-correlation across multiple sensors to construct
more complex labels). 

\subsection{Probabilistic Parameterized Metric Temporal Logic}

\label{sec:ppmtl} 
\begin{figure}[h!]
\includegraphics[width=1\linewidth]{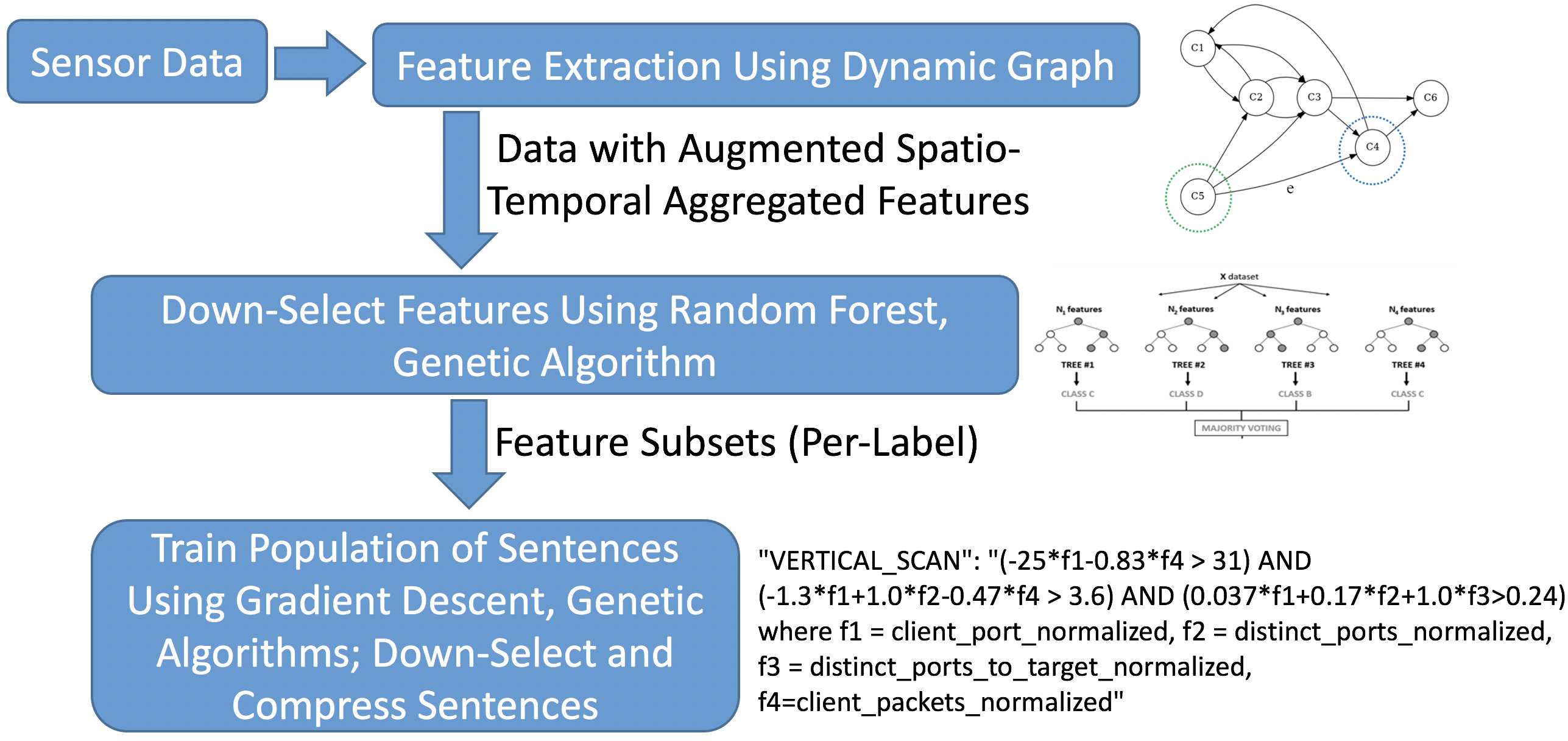} \caption{Human Readable Architecture}
\label{fig:Explainability Architecture} 
\end{figure}

The ppMTL module (Figure~\ref{fig:Explainability Architecture})
in ESAFE can run in two modes: a light-weight mode in which the features
relevant to a classification of a particular label are identified
and a full mode in which additionally a human-readable representation
of the decision logic is generated as a set of linear inequalities
in terms of the identified features. The ppMTL module in ESAFE plays
a complementary role to the GME module. While the GME addresses at-scale
classification from input data, the ppMTL module addresses the generation
of human-readable annotations. Hence, in one configuration, the ppMTL
module is only fed the rows that GME marks as non-normal, i.e., the
rows that the GME associates with at least one label. For these rows,
the ppMTL module then generates human-readable annotations. Since
the ppMTL module also internally generates per-class likelihoods for
input rows as part of the generation of human-readable annotations,
the other configuration possibility is to feed all input data rows
to ppMTL and then probabilistically fuse the outputs of GME and ppMTL
using, for example, a Bayesian per-class ensemble classifier. The
choice between these two configurations is primarily one of scale
with the former being more computationally efficient to address very
large amounts of data. In either configuration, the main role of ppMTL
is to provide greater visibility into the system for the human analyst
as compared to raw probability vectors indicating likelihoods of different
labels and to instead be able to provide an insight into why a given
label fired. 

\subsubsection{Spatio-Temporal Feature Extraction Using a Dynamic Graph}

To facilitate representation of decision logic in terms of human-interpretable
quantities, a spatio-temporal aggregation approach is utilized to
construct features that summarize various aspects of network activity
from the viewpoints of specific nodes (IP addresses) in the network.
This is crucial since the underlying ``reasons'' for why a particular
row in sensor data is classified as a particular label could be related
to events represented in multiple spatially/temporally separated rows
in the sensor data. Hence, simply presenting raw sensor data does
not provide much insight into why a particular label fired; instead,
it is crucial to highlight the aggregates among the rows that together
explain why the label fired. In the context of network log data, spatio-temporal
aggregations include, for example, quantities such as statistics of
incoming and outgoing packets and bytes at each node and between pairs
of nodes (e.g., total, max, average, number of distinct ports), numbers
of distinct nodes with which a node has communicated over a sliding
time window, numbers of distinct ports over which a node has communicated
over a sliding time window, etc. A dynamic graph structure (with graph
nodes being network nodes represented as IP addresses and graph edges
being communications between network nodes) is used to efficiently
ingest and process the streaming sensor data. The dynamic graph is
iteratively constructed from the streaming sensor data and provides
a framework to instantiate new nodes and edges depending on observed
network activity, maintain a sliding time window memory, dynamically
remove nodes and edges based on configurable forgetting rules, and
enable computation of spatio-temporal aggregation based features as
queries answerable by nodes and edges in the graph (by consulting
adjacent/nearby nodes and edges as required for specific spatio-temporal
aggregations).

\subsubsection{Feature Down-Selection Using Random Forest and Genetic Algorithms}

To determine the subset of features (including features present in
original sensor data and features constructed using spatio-temporal
aggregations) that are most relevant for classification of each label,
a random forest classifier is used as an efficient per-class feature
utility estimator and a genetic algorithm is used as a search algorithm
in the space of possible feature subsets. While other classification
algorithms could also be used instead for this purpose, a random forest
classifier has the advantage that it intrinsically generates (as part
of training) feature utility/relevance estimates (feature ``importances''),
which are directly provided by highly optimized off-the-shelf implementations,
and is therefore used for feature utility estimation in this implementation.
For each label, starting from a randomly initialized population of
candidate solutions (choices of subsets of features), each candidate
solution is evaluated to form two metrics: firstly, the classification
$F_{1}$ score on a test dataset when the classifier for that label
is trained using only the subset of features in that candidate solution;
secondly, the utility of each feature within the candidate solution
as estimated by the random forest classifier during training. The
fitness of a candidate solution is defined as a composite of the classification
$F_{1}$ score and a penalty based on the number of features appearing
in that candidate solution (with a pre-specified constraint on max
number of features; e.g., 5 in our experiments). Genetic operators
for random mutations and cross-overs are defined to favor picking
candidate solutions of higher fitness and features therein of higher
feature utilities. The subset of features identified as most relevant
for each label is used to learn a set of human-readable sentences
as discussed below that characterizes the classification decision
logic for that label.

\subsubsection{Learning of Human-Readable Sentences}

For each label, based on the subset of relevant features identified
as discussed above, a set of human-readable sentences (with each sentence
comprised of an AND-combination of a set of linear inequality based
conditions, each of which is called a ``word'' in the sentence)
is learned using a combination of stochastic gradient descent and
genetic algorithms~(\citet{KSK21}). This component of ppMTL is enabled only when ppMTL
is run in ``full'' mode and is not enabled in the ``light-weight''
mode in which the feature down-selection discussed above provides
the required identification of the subset of features. The learning
of the human-readable sentences addresses two problems: learning of
which features to use in each inequality (as defined by a binary mask
over the set of all features) and learning of coefficients in the
inequalities. The union of the set of all sentences learned for a
label characterizes (via an OR-combination) the volume in feature
space corresponding to the label. Discrete-valued features (i.e.,
features that can take only a relatively small number of possible
values, e.g., Boolean-valued features) are encoded into continuous-valued
proxy features based on, for example, relative frequencies of the
appearance of the possible values for the features.

To enable gradient-based learning of the coefficient variables, a
differentiable relaxation of the OR-of-AND structure of the set of
sentences is used based on sigmoid functions and replacement of Boolean
operators by product (for AND) and max (for OR) operators. The loss
function for training is defined as a combination of the binary cross-entropy,
a penalty term for false positives, and a regularizer penalizing the
magnitudes of the coefficient variables and number of non-zero entries
in the mask variables. In parallel with the learning of the coefficient
variables using stochastic gradient descent (e.g., every some pre-defined
number of epochs of gradient-based updates), the mask variables are
updated using a genetic algorithm based on a fitness model for sentences.
The fitness of a sentence is defined as a combination of the classification
performance of the sentence as measured by the rates of true positives
and false positives and the sentence complexity (based on the number
of features appearing in the sentences). The set of sentences is randomly
initialized, fine-tuned using gradient-based updates, and evolved
using randomized mutation and resampling operators based on the computed
fitness values of the sentences.

After the sets of sentences are learned for each label, these sets
are pruned using a sentence down-selection algorithm to determine
the subsets of sentences that are adequate to represent the learned
model and each retained sentence is compressed to remove redundancies
in the constituent inequalities and determine a more compact representation
of the sentence. Once separate models are learned for each anomaly/threat
type, inter-label correlations (i.e., between different anomaly/threat
types) can be integrated based on any known inter-dependencies between
labels (e.g., due to semantic interrelationships between labels) and
learning of a meta-classifier model to map the activation outputs
of the classifiers for each of the labels into the final outputs for
the labels.

\subsubsection{Human Readable Output}

Based on the learned feature subsets and sentences as discussed above,
the ppMTL module can identify, during inference, the features and
linear inequality based combinations thereof for each specific row
in sensor data classified as a particular label. The ppMTL module
can also identify the composite of all relevant features and linear
inequality conditions for all data points classified as that label
(i.e., the union of human-readable annotations which serves as a description
of the overall decision logic for that label). The ppMTL module's
human-readable annotation outputs for each row classified as a particular
label are passed to an ancillary database that is queryable by an
analyst or post-processing module for further downstream use. While
multiple sentences could fire for a particular row, ppMTL automatically
picks the best sentence based on sentence compactness and classification
performance metrics. The selected sentences provide context-aware
explanations for the classification by identifying the specific set
of conditions that were satisfied by each data point marked as a particular
label.

\section{System Architecture}

\label{sec:system_architecture} The ESAFE system is a hierarchical
reasoning system comprised of layers of data storage, collection and
preprocessing, machine learning based classification using the GME
and ppMTL, and finally postprocessing over the final output before
transforming it into the desired UI.

\subsection{Pre-Processing}

The ESAFE system is designed to ingest a variety of data types from
a variety of data sources. Furthermore, it is designed to be extensible,
so that new data types and data sources can be added without re-implementing
the machine learning and clustering components of the system.

In order to support such a wide variety of data types and sources,
as well as to ensure extensibility, the data must be pre-processed
into a common format that can be consumed by subsequent components.

The Pre-Processing stage comprises the following steps: 
\begin{enumerate}
\item Identify available data types. 
\item Define a schema for each data type. 
\item Collect the raw data from one or more data sources. 
\item Convert the raw data into one or more standardized formats that correspond
to the schema for that data type. 
\item (Optional) Label the data. 
\end{enumerate}
Each of these steps is discussed in further detail below.

\subsubsection{Identifying Available Data Types}

The ESAFE system is designed as a general framework for processing
cyber-relevant data from a variety of data sources. ESAFE does not
perform data collection on its own; instead, it relies on data that
has been collected by other tools (aka ``data sources''), such as
Bro/Zeek, NetFlow, Windows Event Logs, firewall logs, etc.

Consequently, one of the first steps is to identify the data types
that the system will process. For our purposes, a ``data type''
is a record with a consistent schema that is generated by a specific
monitoring tool. The following are examples of some data types that
are supported by the ESAFE system: 
\begin{enumerate}
\item Bro/Zeek Conn logs 
\item Bro/Zeek SSH logs 
\item NetFlow records 
\item Active Directory event logs 
\item Web Firewall event logs 
\end{enumerate}

\subsubsection{Collecting Raw Data}

A particular concern at this step is how to retrieve the data at scale.
Some data types, such as Bro/Zeek Conn logs and NetFlow logs, are
voluminous, so care must be taken to ensure that the system can retrieve
the raw data for the desired time range in a reasonable period of
time. In one environment, this was achieved by designing the code
to take advantage of the compute resources that were available to
individual virtual machines (32 vCPUs, up to 256GB of RAM, 2TB of
disk, etc.). In another environment, this was achieved by leveraging
the available high-performance computing (HPC) resources; generally,
this meant breaking the retrieval task up into dozens or hundreds
of very small tasks that could all be run independently in parallel.

There is no single approach that will be optimal for all deployment
environments. Regardless of how the raw data is ultimately collected,
it must then be converted into a format which corresponds to the appropriate
schema for the data type, and then exported into one or more common
formats that can be leveraged by downstream components.

During development, it became clear that the labeling component and
the cluster hunting component relied heavily on the time-ordering
of data. Consequently, the performance of downstream components can
be improved (at times dramatically) by ensuring that when data is
written out, it is written out sorted by time.

\subsubsection{Labeling the Data}

Once the data has been exported to a standard format for processing,
it can optionally be labeled. Labeling rules can vary dramatically
in terms of the complexity of their implementation. When developing
the ESAFE system, we considered the following types of labels: 
\begin{enumerate}
\item A label which operates on a single row of data. This is the most scalable
type of label. 
\item A label which operates on a relatively small window of data, such
as 100 to 1000 rows, from the same file. This type of rule generally
scales well. 
\item A label which operates on a ``large'' window of data of the same
data type, where ``large'' will vary based on data type. Typically,
anything that will require tens of thousands of rows (or more) split
across multiple files would fit in this category. In general, this
type of rule is more difficult to scale. 
\item A label which operates on multiple data types at once. In this study,
we did not implement labels of this type. 
\end{enumerate}
These are discussed in greater detail below. \\

\subsubsection{Labeling Single Rows of Data}

The easiest and most scalable types of labels to apply are labels
that can be derived from examining one row of data at a time. For
labels such as these, the labeling task can be broken down into hundreds
of smaller labeling tasks which can then be submitted as a batch job
on a HPC system. These rules are also easier and faster to implement,
as well as less prone to errors (bugs), because they do not need to
maintain state.

However, these labels are also generally the least ``interesting''
from an analyst's perspective.

\subsubsection{Labeling Based on Small Windows}

Labels that are applied based on ``small windows'' of data can provide
much of the scalability benefits of single-row-based labels while
allowing for more complexity (and analyst utility). Typically, windows
that are defined in terms of a fixed number of rows (e.g., 1000 rows)
are the most scalable and easiest to implement; windows that are defined
in terms of a fixed period of time (e.g., 10 seconds) can be more
difficult to implement, because for some data types, such as Bro/Zeek
Conn logs, 10 seconds of data could represent hundreds of thousands
of rows; furthermore, when windows are defined in terms of time, there
is often a greater likelihood that the data will be split across multiple
files.

Nevertheless, some labels may require time-based windows due to their
nature. As discussed before, the performance of such labels will typically
benefit if the data was exported in time-series order (i.e., sorted
by time).

\subsubsection{Labeling Based on Large Windows}

As a general rule, labeling based on ``large windows'' (either millions
of rows or hours of data) will require a more sophisticated back-end
data store. Typically, in order to apply such labels in a timely manner,
the system will need to leverage a data store that supports queries.
Scalability in labeling when using large data windows remains an open
topic of research.

\subsection{Cluster Hunting and Post-Processing}

Once the data has been labeled (enriched) and filtered by the GME
and the ppMTL, numerous data science and analytical techniques can
be applied to search for suspicious and malicious behavior during
cluster hunting: 
\begin{enumerate}
\item Describe a type of malicious behavior that is to be searched 
for in the data. 
\item Identify artifacts that would be left in the available logs from such
behavior. 
\item Apply a label indicating each such artifact. 
\item Define a ``cluster'' of such labels which indicate that behavior.
Minimally, a cluster would be a group of labels that appear in a specific
order within a particular time window. Typically, clusters will have
additional constraints, such as the labels being applied to the same
source or destination IP addresses, or the same TCP or UDP ports,
etc. 
\item Search for all defined clusters in the labeled data. 
\item Report each detected cluster via an appropriate mechanism, such as
a report, a log entry, a ticket submitted to a ticketing system (such
as Redmine), etc. 
\end{enumerate}
Steps 1-4 were the focus of Section 3
while steps 5-6 are explored in more detail below.

\subsubsection{Label Artifacts}

Once 3-5 artifacts have been identified that are associated with a
class of malicious behavior, it will need to be ensured that each artifact
label exists in the pre-processing of data. If such labels do not
already exist, they will need to be added. Adding new labels to the
system can be time-consuming, since it requires re-labeling training
and test data, and then re-training the GME and ppMTL. However, one
of the strengths of the Cluster Hunting approach is that the same
label can appear in any number of clusters; so it will not necessarily be
required to define a new label whenever a new cluster is added to the system.
For example, two different clusters might both include an underlying
label such as HTTP\_REDIRECT.

\subsubsection{Search for Clusters}

If a data store is available which supports a sophisticated query
language, such as SQL, then it may be possible to formulate the cluster
logic as a single query to the system. Assuming that a suitable, query-able
data store is not available, the labeled data
files will need to be searched. During our experiments, we also found that removing rows from
the data set which did not have labels that occurred in at least one
cluster could substantially reduce the overall volume of data that
needed to be searched. In one case, we were able to reduce the overall
volume of data for a particular data type by half. This is often the
case with a ``Normal'' label\textemdash{} i.e., a
label which represents a row that has no other label. Since the ``Normal''
label never occurred in any of our cluster definitions, we were able
to reduce the overall search space by half simply by eliminating rows
that had been labeled ``Normal''.

\subsubsection{Report Detected Clusters}

The appropriate mechanism to report detected clusters will vary based
on the deployment scenario. Some analysts will simply want a report,
such as a text file, to describe the detected cluster. Other scenarios
may require exporting the cluster in a machine-consumable format,
such as JSON file. Other scenarios may involve publishing the detected
cluster to a message channel, such as Apache Kafka, for use by downstream
software. And still other workflows may require posting the detected
cluster as an issue to an issue tracker or ticketing system, such
as Redmine. 
\begin{figure}[h!]
\includegraphics[width=1\linewidth]{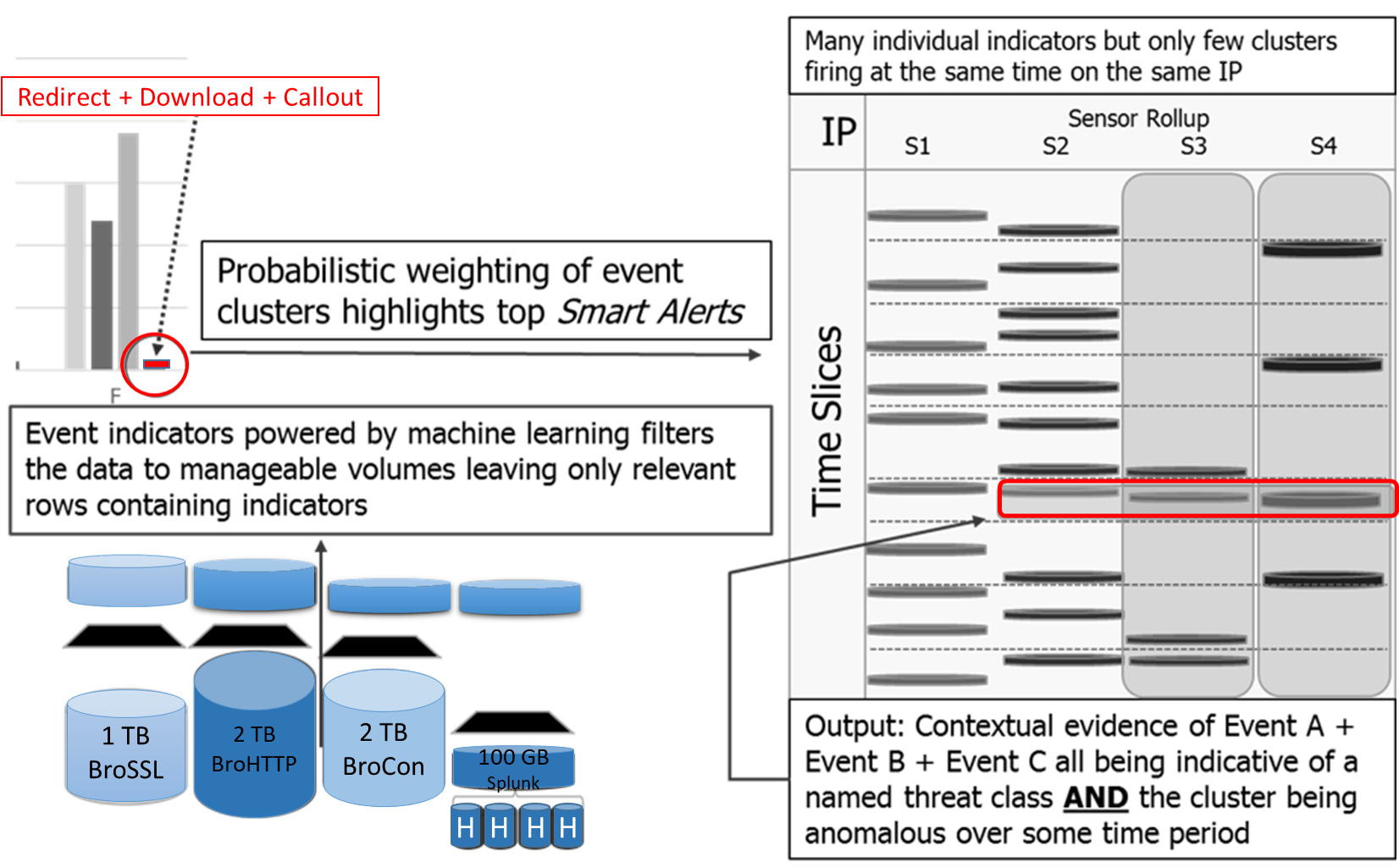} \caption{Smart Alert Flow}
\label{fig:Smart Alerts} 
\end{figure}

In general, the cluster hunting system should be designed in such
a way as to support exporting detected clusters to a variety of export
formats. As with the \emph{Collecting Raw Data} step in the \emph{Pre-Processing}
stage, this is likely to be a part of the system that requires a nontrivial
degree of environment-specific (\citet{2009IEEE}) development. (\citet{Stide}).
\\

\section{Lessons Learned }

\label{sec:lessons_learned} We have four key lessons learned we wish
to highlight: 
\begin{enumerate}
\item We must know when we need more data and where it should be collected.
This follows ``The essence of the CHASE program is how do you
get the right data from the right device at the right time in order
to really bolster our security in our networks.'' (\citet{DefenseSystems}). 
\item Logs may not contain all the data needed to make a determination between
conclusions when there are two possible explanations for our cluster
of elemental rules alerting. 
\item There is no single one \emph{best} labeling strategy and we need a
moderated approach where we have enough elemental rules to have clusters
which are explainable to the user and able to find new attacks. 
\item A short time window is not sufficient to alert, any threat that takes
longer than the time window simply will not fire enough elemental
labels to make a cluster we can make a determination on. We did by
hand cover larger time windows to adapt to this limitation. 
\end{enumerate}

\section{Open Challenges }

\label{sec:open_challenges} Inconclusive ESAFE reports is our greatest
current short fall. We have on numerous occasions been left with the
conclusion that there are two possible explanations for our cluster
of elemental rules alerting (e.g., have we observed a watering hole
attack or has there been a redirect to a previously unseen server?
have we observed a DDOS attack or is there a Bro logging error/misconfiguration?).
False positives where alerts are generated since our cluster label(s)
have hit but with inconclusive supporting data result in the Analyst
having to spend a large amount of time to investigate the flagged
anomalies. This is a very cost intensive process and quickly leads
to diminishing returns for the SOC Analyst. Anomalies, particularly
in network traffic, do not necessarily equate to malicious behavior.
The noise floor and the flux in user behavior both contribute to this
effect. This opens the question: was information available sufficient
to determine which of two hypotheses is correct? In cases of ambiguity,
we must infer the data we need to resolve the ambiguity and collect
such data. Waiting for the data to make a decision, benign or malicious,
to arrive is not necessarily a correct approach for two main reasons: 
\begin{enumerate}
\item The data we want may never be transmitted. 
\item Attackers often run campaigns over long periods of time to avoid detection.
Since attacks would take a long period of time to manifest, it means
the number of hypotheses can grow to an unmanageable scale forcing
us to stop tracking. In either case, many attacks can be missed. 
\end{enumerate}
For future work, we recommend two main avenues: firstly, the category
of suspicion should be adopted, in order to bridge the semantic and
information gap; secondly, a study of the information loss from raw
data (e.g., pcap) to logs (e.g., bro\_ssl\_logs) needs to be performed.

\subsection{Study Information Loss from raw data to logs }

All of our elemental labels reason over logs and when ESAFE reports
back, the highest degree of fidelity of evidence that can be provided
is row(s) of data with columns of interest included. Currently, SOCs
investigate alerts and pivot as necessary to obtain supporting information.
The analyst knows what information is relevant and from what source
or sources to obtain it. Future work should apply information theoretic
approaches to quantify what information is lost at each stage of collection
and what information is needed to make a determination for each elemental
rule and cluster label. Logs from sensors are not intended to be complete
forensic archives of network traffic. In fact, they are by definition
and construction a summary of the expected behavior. \\
 As a motivating example, we see the need to cluster information-aware
rules: Let us consider a Bro\_SSL\_Log and Artifacts from the set
\char`\"{}TTP\char`\"{} \_(TLS\_encrypt\_c2), where the JA3 hash is
present in an SSL Log, the client\_hello information has been hashed
(via MD5) and thus extracting similarities from the artifacts is not
possible from the log. Since real browsers support $\sim$40 encryption
methods, logging of all client\_hello strings would collect too much
data resulting in log storage problems.

In accordance with the CHASE model, information-aware rules are used
for finding similarities of client\_hello strings to known malware
samples. We would need to request this data (via pcap) iff a threshold
for suspicion is met for a specific server; this data would not be
stored after comparison. This work should be approached by describing
what information is captured by each log and what information is present
but not captured. Elemental rules should clearly include in their
automated report what information is needed to make a determination,
but was not present (\citet{SecureML}).

\subsection{Suspicion}

Even with high fidelity in the information that is present in log(s)
and that is available from raw data, many alerts will not be resolved
in a rolling time window. The concept of a system being \emph{suspicious}
should be introduced. When sufficient data (e.g. logs, raw) to resolve
if an alert is malicious is not available, such suspicions must still
be presented to the analyst for their interpretation and not simply
dismissed.

There are some mitigating steps ESAFE should explore, such as additional
targeted information via raw collection. This should be done in two
ways; with respect to the systems involved in the alert, more data
requested by ESAFE and with respect to the network the meta-statistics
of alert clusters should also be tracked. ESAFE should also track
when new elemental labels appear in clusters they have not been seen
in previously, or when more data is requested by ESAFE than had previously
been needed for a cluster alert. In other words, ESAFE must track
how cluster alerts are currently being used and how they are changing.

\bibliographystyle{plainnat}
\bibliography{references}
 \global\long\def\refname{References}

\end{document}